\documentclass[%
 reprint,
superscriptaddress,
showpacs,preprintnumbers,
 amsmath,amssymb,
 aps,
]{revtex4-1}
\usepackage{soul}
\usepackage{ulem}
\usepackage{graphicx}
\usepackage{dcolumn}
\usepackage{bm}
\usepackage{color}
\newcommand\redout{\bgroup\markoverwith{\textcolor{red}{\rule[.5ex]{2pt}{0.4pt}}}\ULon}

\usepackage[colorlinks=true,citecolor=blue]{hyperref}
\hypersetup{colorlinks=true,citecolor=blue,linkcolor=red,urlcolor=blue}

\begin{document}
\date{\today}
\title{Giant magnetoresistance and anomalous transport in phosphorene-based multilayers with noncollinear magnetization}

\author{Moslem Zare}
\affiliation {School of Physics, Institute for Research in Fundamental Sciences (IPM), Tehran 19395-5531, Iran}
\author{Leyla Majidi}
\affiliation{School of Nano Science, Institute for Research in Fundamental Sciences (IPM), Tehran 19395-5531, Iran}
\author{Reza Asgari}
\affiliation{School of Physics, Institute for Research in Fundamental Sciences (IPM), Tehran 19395-5531, Iran}
\affiliation{School of Nano Science, Institute for Research in Fundamental Sciences (IPM), Tehran 19395-5531, Iran}

\date{\today}

\begin{abstract}
We theoretically investigate the unusual features of the magnetotransport in a monolayer phosphorene ferromagnetic/normal/ferromagnetic (F/N/F) hybrid structure. We find that the charge conductance can feature a minimum at parallel (P) configuration and a maximum near the antiparallel (AP) configuration of magnetization in the F/N/F structure with $n$-doped F and $p$-doped N regions and also a finite conductance in the AP configuration with the N region of $n$-type doping. In particular, the proposed structure exhibits giant magnetoresistance, which can be tuned to unity. This perfect switching is found to show strong robustness with respect to increasing the contact length and tuning the chemical potential of the N region with a gate voltage. We also explore the oscillatory behavior of the charge conductance or magnetoresistance in terms of the size of the N region. We further demonstrate the penetration of the spin-transfer torque into the right F region and show that, unlike graphene structure, the spin-transfer torque is very sensitive to the chemical potential of the N region as well as the exchange field of the F region.
\end{abstract}

\maketitle

\section{Introduction}
Recently, a new type of material consisting of a black phosphorus monolayer (phosphorene) has emerged as a viable candidate in the field of two-dimensional (2D) materials~\cite{Li14,Liu14, andre}. Unlike graphene which has a planar structure, phosphorene forms a puckered honeycomb structure, owing to an $sp^3$ hybridization with a high in-plane structure anisotropy which has been affected in the optical and transport properties of phosphorene~\cite{Liu15,Das14,Kamalakar15,Lu14,Fei14,Wang1}. Charge-carrier mobilities are very high at room temperature $\sim 10^3$ cm$^2$ V$^{-1}$ S$^{-1}$~\cite{Li14} and they exhibit a strongly anisotropic behavior in phosphorene-based field effect transistor with a high on/off ratio of $10^4$~\cite{Liu14}. The band structure of  few-layer black phosphorus has a direct gap, spanning a wide range in the visible spectrum from $0.8$ to $2$ eV~\cite{Du10,Tran14,cakir14,Liang14,Rodin14}. The edge magnetism has been explored in different edge cutting directions~\cite{Zhu14,Du15,Yang16,Farooq15,Farooq16}. The introduction of $3d$ transition-metal atoms induces magnetism, where the magnitude of the magnetic moment depends on the metal species and the result can be tuned by the applied strain~\cite{Hu15,Sui15,Seixas15}. Also, it is found that a spin-polarized state appears in monolayer black phosphorus by nonmagnetic impurity doping~\cite{khan15}. These properties make phosphorene a quite interesting 2D material for next-generation electronic and spintronic devices based on the spin degree of freedom, which are almost dissipationless unlike those based on the charge degree of freedom~\cite{fabian}.

The conservation of angular momenta between itinerant electrons and localized magnetization in magnetic materials leads to the fascinating concept of spin-transfer torque; the spin angular momentum of the electrons can be transferred to the magnetization via their mutual exchange coupling which enables us to derive the dynamics of magnetization by charge current. Sloncwezski and Berger were the first to theorize about the existence of this phenomenon~\cite{Berger,Slonczewski}. It is found to be important in all known magnetic materials and present in a variety of material structures and device geometries composed of magnetic-nonmagnetic multilayers such as magnetic tunnel junctions, spin valves, point contacts, nanopillars, and nanowires~\cite{Brataas}. The spin current flowing into the magnetic region exerts a torque on the magnetization, if the current polarization direction is non-collinear to the local magnetization in the magnetic material. This spin-transfer torque effect can cause magnetization switching for sufficiently large currents without the need for an external field. The switching aspect provides a unique opportunity to create fast-switching spin-transfer torque magnetic random access memories (STT-MRAM)~\cite{Parkin}, with the advantage of low power consumption and better scalability over conventional MRAMs which use magnetic fields to flip the magnetization. The spin-torque oscillator is another prominent example that serves as a field generator for microwave-assisted recording of hard-disk drives~\cite{Zhu08}. Also, the spin-transfer torque leads to the spin-torque diode effect in magnetic tunnel junctions~\cite{Tulapurkar}. Therefore, this new discovery in condensed matter and material physics has expanded the means available to manipulate the magnetization of magnetic materials and as a result has accelerated technological development of high-performance and high density magnetic storage devices~\cite{fabian,Fert,Brataas,Bauer,linder15}. Another important quantity in this field is the tunneling magnetoresistance which occurs at ferromagnetic/normal/ferromagnetic (F/N/F) hybrid structures. The resistance of the junction is different for parallel (P) and anti-parallel (AP) magnetiozation configurations and it can be experimentally measured~\cite{exp1,exp2}. Spin-polarized resonant tunneling has been intensely studied~\cite{Glazov,Zheng,Ertler} because of its potential for applications in spin-transfer torque switching enhancement~\cite{Chatterji}, cavity polaritons using the resonant tunneling diode~\cite{Nguyen}, and in the spin-dependent resonant tunneling devices for tuning the tunneling magnetoresistance~\cite{Yuasa, Petukhov}.

In the present work, we investigate the charge and spin transport characteristics of the monolayer phosphorene F/N/F hybrid structure in which the two F regions with noncollinear magnetization are connected through the normal segment of length $L$. We model our system within the scattering matrix formalism~\cite{datta}. A similar study has been utilized in graphene~\cite{fnfgraphene1,fnfgraphene2} and silicene~\cite{fnfsilicene}. Within the scattering formalism, we find that the transmission probability of an incoming electron from the spin ${\sigma}$ band of the left F region of the F/N/F structure with P configuration of magnetization has a decreasing behavior with respect to an increase of the incidence angle, while a peak structure appears for an incoming $\bar{\sigma}$-band electron in the proposed structure with small length $L$. We demonstrate the appearance of a peak structure with perfect transmission for the ${\sigma}$-band electron together with additional peaks in the transmission probability of the $\bar{\sigma}$-band electron, by increasing the contact length $L$. However, both the incoming electrons from the $\sigma$  and $\bar{\sigma}$ bands to the F/N/F structure with AP alignment of magnetization vectors have decreased transmission probabilities with the increase of the angle of incidence for small contact lengths and peak structures for large contact lengths.

The application of a local gate voltage to the large-length N region of the F/N/F structure leads to the amplification of transmission probability at defined angles of incidence and the attenuation of the transmission at other traveling modes in the proposed structure with the $p$-doped N region. The $n$- or $p$-type doping is the process of enhancing the magnitude of the Fermi energy in the conduction or valence band of the material, tuning the electron or hole charge carriers. Then, we evaluate how it is possible to control the magnitude of the charge conductance by means of a local gate voltage in the N region, and manipulating the magnetization directions of two F regions. Depending on the chemical potential of the F region $\mu_{\rm F}$, the charge conductance can be suppressed away from the AP alignment of magnetization and finite at AP alignment in the n-type doped F/N/F structure. In particular, it displays a maximum near the AP configuration and a minimum at the P configuration in  the corresponding structure with p-doped N region. We further demonstrate that the proposed spin valve structure exhibits a remarkably large magnetoresistance in analogy to the giant magnetoresistance in magnetic materials~\cite{exp1}, which can be tuned to unity for determined ranges of $\mu_{\rm F}$. More importantly, we find that the perfect spin valve effect is preserved for large lengths of the N region and shows strong robustness with respect to increasing the chemical potential $\mu_{\rm N}$ with the local gate voltage. In addition, we compute the equilibrium magnetic torque exerted on the magnetic order parameter of the right F region along the $z$ and $y$ directions and demonstrate the penetration of the torque into the F region. We show that the sign change of the torque can be realized by means of the gate voltage in the N region and tuning the exchange field of the F region, in addition to its magnitude changes.

This paper is organized as follows. In Sec. \ref{sec:level1}, we introduce the model and establish the theoretical framework which is used to calculate the transmission probabilities, charge conductance, spin-current density, and the spin-transfer torque. In Sec. \ref{sec:level2}, we present and describe our numerical results for the proposed phosphorene-based F/N/F structure. Finally, our conclusions are summarized in Sec. \ref{sec:level3}.

\section{Model and basic equations}\label{sec:level1}
\begin{figure}[]
\begin{center}
\includegraphics[width=3.4in]{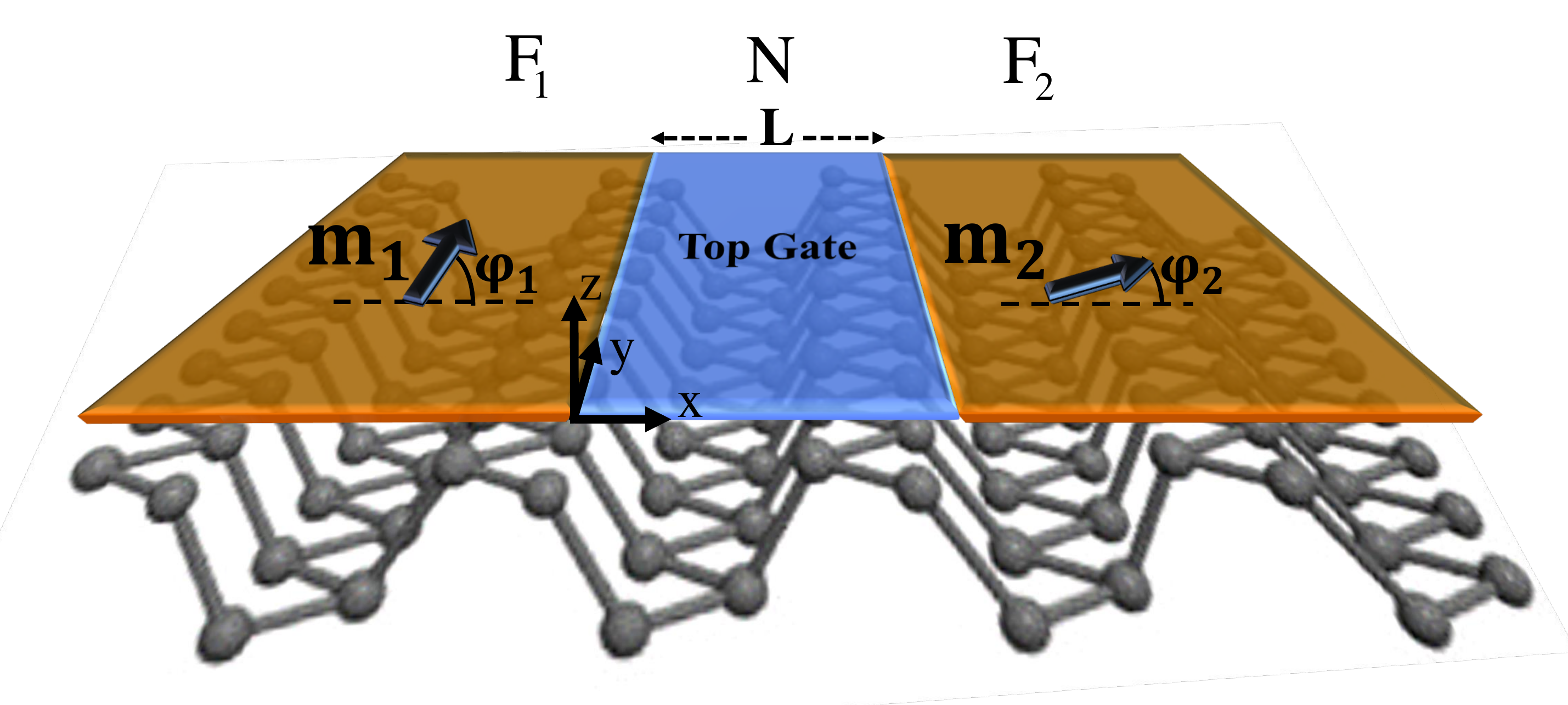}
\end{center}
\caption{\label{Fig:1} Schematic illustration of the monolayer phosphorene F/N/F junction: The two F regions with the magnetization vectors $\bm{m}_i$ are coupled through the N region located at $0<x<L$.}
\end{figure}

\begin{figure}[]
\begin{center}
\includegraphics[width=3.4in]{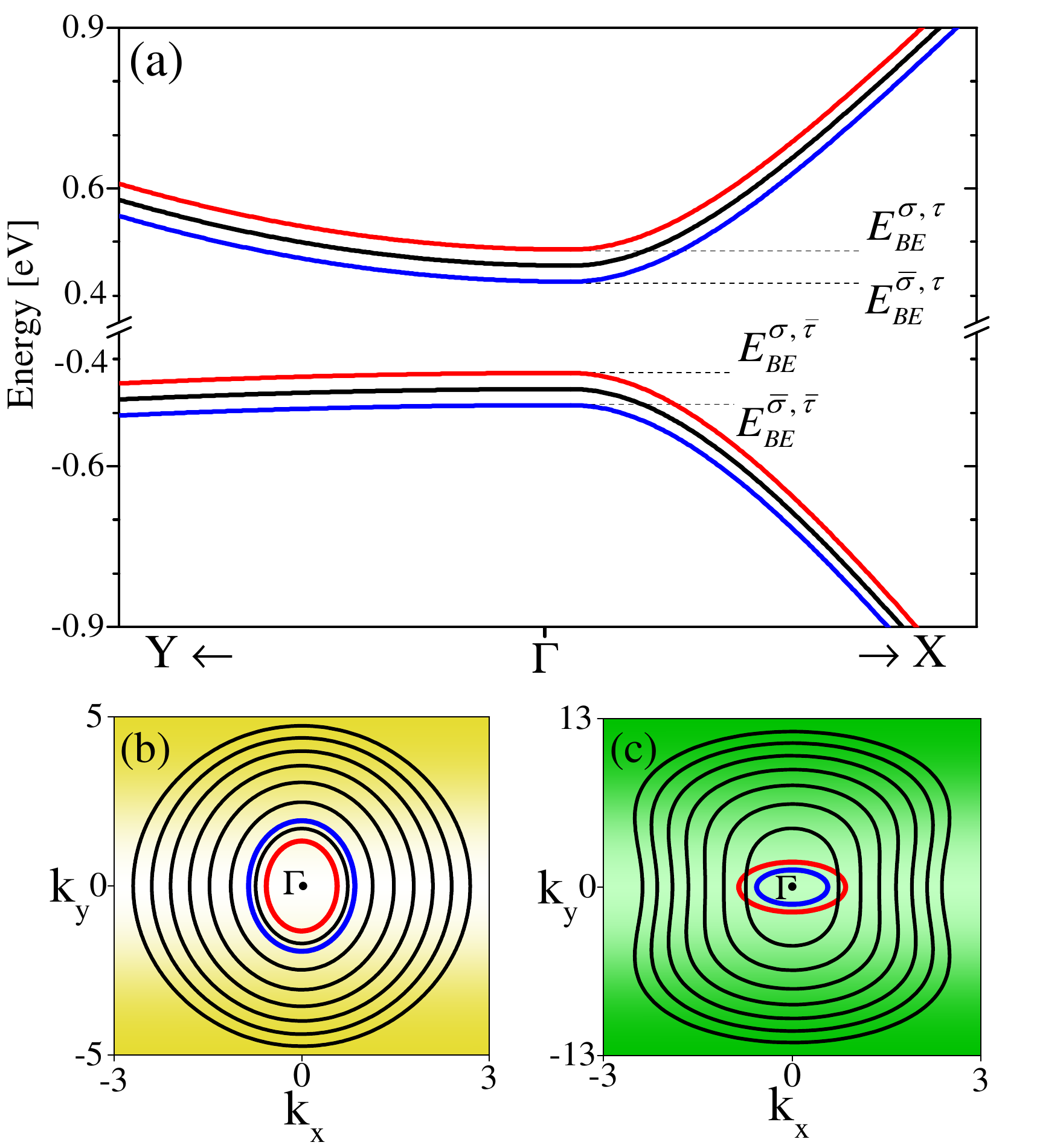}
\end{center}
\caption{\label{Fig:2} (a) The dispersion relation in momentum space of the monolayer phosphorene along the Y-$\Gamma$-X direction in the Brillouin zone. The black lines correspond to the N phosphorene, while the red (blue) line corresponds to the $\sigma$ ($\bar{\sigma}$) band of the F phosphorene with the magnetization $m=0.03$ eV. (b),(c) Isofrequency contour surfaces in the $k$ space of $n$-type and $p$-type doped phosphorene layers, respectively, for $E=0.4557$ eV and $E=-0.4557$ eV, with the step of $0.1$ eV. Red (blue) isofrequency contour surface corresponds to the $\sigma$ ($\bar{\sigma}$) band of the F phosphorene at $E=\mu_F\pm m$, with the chemical potential $\mu_F=0.54$ eV. The energies $E^{{\sigma}, \tau(\bar{\tau})}_{BE}$ and $E^{\bar{\sigma}, \tau(\bar{\tau})}_{BE}$ define the conduction (valance) band edges for the two spin subbands of F phosphorene ($\sigma=1$, $\bar{\sigma}=-\sigma$) with $\tau=1$ ($\bar{\tau}=-\tau$). Note that the zero-point energy in the band structure of N phosphorene is shifted to the middle of its energy band gap.}
\end{figure}
Let us consider a wide monolayer phosphorene F/N/F hybrid structure in the $x-y$ plane, where the N region of length $L$ couples two F regions with the magnetization vectors $\bm{m}_1=m (\cos\phi_1,\sin\phi_1)$ and $\bm{m}_2=m (\cos\phi_2,\sin\phi_2)$, as sketched in Fig. \ref{Fig:1}. The interface between the N and F regions are located at $x=0$ and $x=L$. It is assumed that the ferromagnetism can be induced by means of the proximity effect when it is placed with a magnetic insulator. Such an F region in graphene can be produced by using an insulating ferromagnetic substrate, or by adding F metals or magnetic impurities on top of the graphene sheet~\cite{Tombros07,Haugen,Swartz12,Wang15,Dugaev06,Uchoa08,Yazyev10}. The direction of the magnetization vector can be controlled by externally applied magnetic field.

The system is translationally invariant along the $y$ direction and thus the y-component of the wave vector $k_y$ does not change and is conserved. The effective low-energy Hamiltonian of a quasiparticle in a monolayer phosphorene, in the presence of a magnetization vector $\bm{m}$ in the F region, has the form
\begin{equation}
\label{H}
{\cal H}={\cal H}_0 \otimes \hat{\sigma}_0+\hat{\tau}_0\otimes \bm{m}\cdot\hat{\bm{\sigma}}-\mu_F
\end{equation}
which acts on a four-component spinor $(\Psi_{c\sigma},\Psi_{c\bar{\sigma}},\Psi_{v\sigma},\Psi_{v\bar{\sigma}})$. The indices $c,v$ respectively label the conduction and valence bands, and $\sigma$, $\bar{\sigma}$ ($\bar{\sigma}=-\sigma$) denote the two spin subbands. The two sets of Pauli matrices, $\hat{\sigma}$ and $\hat{\tau}_0$, act on the real spin and the conduction and valence band pseudospin degrees of freedom, correspondingly, and $\mu_F$ is the chemical potential of the F region. The two-dimensional $\bm{k}\cdot\bm{p}$ model Hamiltonian ${\cal H}_0$ in the subspace of the conduction and valence bands can be described by
\begin{equation}\label{H0}
{\cal H}_{0}=
\begin{pmatrix}
E_c+\eta_c k_x^2+\nu_c k_y^2 &\gamma k_x& \\ \gamma k_x& E_v-\eta_v k_x^2-\nu_v k_y^2
 \end{pmatrix},
\end{equation}
where the energies of the conduction and valence band edges, $E_c=-0.3797$ eV and $E_v = -1.2912$ eV, and the coefficients $\eta_c=0.008187$, $\eta_v=0.038068$, $\nu_c= 0.030726$, $\nu_v= 0.004849$ in units of eV nm$^2$, and $\gamma= 0.48$ in units of eV nm, are obtained from the first-principles simulations based on the density-functional theory (DFT)~\cite{Zare, asgari}.

Figure \ref{Fig:2}(a) shows the anisotropic band energy dispersion of the monolayer phosphorene along the $Y-\Gamma-X$ direction in the Brillouin zone, where the zero-point energy is shifted to the middle of the band gap. The isofrequency contour surface in the $k$ space of the $n$- ($p$-) doped monolayer phosphorene demonstrates the elliptic shape of the Fermi surfaces especially at low charge densities, as seen in Figs. \ref{Fig:2}(b) and \ref{Fig:2}(c). The black lines correspond to the N phosphorene, while the red (blue) line corresponds to the $\sigma$ ($\bar{\sigma}$) band of the F phosphorene.

Since it is not easy to control the angle of incident charge carriers, we usually compute the conductance by integrating over the possible angles of incidence. To evaluate the charge conductance of the proposed structure at zero temperature, we use the Landauer formula~\cite{landauer}
\begin{equation}
\label{conductance}
G (E) = \frac{g_0 W}{2\pi}\sum_{{\sigma}_i,{\sigma}_j=\sigma,\bar{\sigma}}\int_{0}^{k'_{{\sigma}_i}(E)} T_{{\sigma}_i{\sigma}_j}(E,k_y)\ dk_y,
\end{equation}
where $T_{{\sigma}_i{\sigma}_j}(E,k_y)$ ($T_{{\sigma}_i{\sigma}_j}=|t_{{\sigma}_i{\sigma}_j}|^2$) describes the transmission probability of an incoming electron from the ${\sigma}_i$ band of the left F region with an energy $E$ to the ${\sigma}_j$ band of the right F region, $g_0=e^2/h$ is the quantum of the conductance, and $k'_{{\sigma}_i}(E)= [(E-E_{c}-{\sigma}_i m)/\nu_{c}]^{1/2}$.

We calculate the transmission amplitudes of the incoming $\sigma$-band electron $t_{\sigma\bar{\sigma}}$ and $t_{\sigma{\sigma}}$, by matching the wave functions of the three regions of the left F, N region, and the right F region (signed by 1, 2, and 3 respectively) at the two interfaces. The total wave functions in the three regions are as follows:
\begin{eqnarray}\label{total wave functions}
\Psi_1&=&\Psi^{F_1+}_{\sigma,\tau}+r_{\sigma\sigma}\Psi^{F_1-}_{\sigma,\tau}+r_{\sigma \bar{\sigma}}\Psi^{F_1-}_{\bar{\sigma},\tau},\\
\Psi_2&=& a\Psi^{N+}_{\sigma,\tau}+b\Psi^{N-}_{\sigma,\tau}+c\Psi^{N+}_{\bar{\sigma},\tau}+d\Psi^{N-}_{\bar{\sigma},\tau},\\
\Psi_3&=&t_{\sigma\sigma}\Psi^{F_2+}_{\sigma,\tau}+t_{\sigma \bar{\sigma}}\Psi^{F_2+}_{\bar{\sigma},\tau},
\end{eqnarray}
where,
\begin{eqnarray} \label{psiF}
\Psi_{\sigma,\tau}^{F_i\pm}&=&
A_{\sigma,\tau}^{F_i} e^{ \pm i\tau k_x^{F_i} x} e^{ik_yy}
\begin{pmatrix}
 \pm\sigma\chi_{\tau}^{F_i} e^{-i\phi_i} \\\pm\chi_{\tau}^{F_i}  \\  \sigma e^{-i\phi_i} \\1
\end{pmatrix},
\end{eqnarray}
and
\begin{eqnarray} \label{psiN1}
\Psi_{\sigma,\tau}^{N\pm}&=&
\sqrt{2}\ A_{\sigma,\tau}^N e^{\pm i\tau k_{x}^N x} e^{ik_yy}
\begin{pmatrix}
\pm\chi_{\tau}^N \\ 0\\1\\0
\end{pmatrix},
\end{eqnarray}

\begin{eqnarray} \label{psiN2}
\Psi_{\bar{\sigma},\tau}^{N\pm}&=&
\sqrt{2}\ A_{\bar{\sigma},\tau}^Ne^{\pm i\tau k_{x}^N x} e^{ik_yy}
\begin{pmatrix}
0 \\ \pm\chi_{\tau}^N\\ 0\\1
\end{pmatrix}.
\end{eqnarray}

They are, respectively, the solutions of Eq. (\ref{H}) for incoming and outgoing electrons of the $F_i$ region with $\bm{m}_i=m(\cos\phi_i,\sin\phi_i)$ and the N region (with $m=0$) at given energies $\varepsilon_{\sigma,\tau}^{F_i}$ and $\varepsilon_{\sigma,\tau}^{N}$, and transverse wave vector $k_y$, with the energy-momentum relation
\begin{widetext}
 \begin{eqnarray}\label{Energy-momentum}
 \varepsilon_{\sigma,\tau}^{N(F_i)}=\frac{1}{2}[{\mathcal{H}_{c}^{N(F_i)}+ \mathcal{H}_{v}^{N(F_i)} +\tau \sqrt{4{(\mathcal{H}_{cv}^{N(F_i)})}^2 + (\mathcal{H}_{c}^{N(F_i)} -\mathcal{H}_{v}^{N(F_i)})^2}}+2\sigma m]-\mu_{N(F_i)}.\nonumber\\
\end{eqnarray}
Here, $\mu_{N(F_i)}$ is the chemical potential of the N($F_i$) region, $\mathcal{H}_c^{N(F_i)}=E_c+\eta_c {(k_x^{N(F_i)})}^2+\nu_c k_y^2$, $\mathcal{H}_v^{N(F_i)}=E_v-\eta_v {(k_x^{N(F_i)})}^2-\nu_v k_y^2$, $\mathcal{H}_{cv}^{N(F_i)}=\gamma k_x^{N(F_i)}$, $\tau=\pm 1$ denotes the conduction (valence) band, $\sigma$ and $\bar{\sigma}$ ($\bar{\sigma}=-\sigma=-1$) refer to the two spin subbands, and $\theta_{N(F_i)}=\arctan(k_y/k_x^{N(F_i)})$ indicates the angle of propagation of the electron. Also, the two propagation directions along the $x$ axis are denoted by $\pm$ in $\Psi_{\sigma,\tau}^{N(F_i)\pm}$. Note again that $m=0$ in the N region.

The propagating electron in the N(F$_i$) region with the velocity
 \begin{equation}
 v_{x,\tau}^{N(F_i)}=k_{x}^{N(F_i)}\left[\eta_{c}-\eta_{v}+\tau\frac {2\gamma^{2}+(H_{c}^{N(F_i)}-H_{v}^{N(F_i)})(\eta_{c}+\eta_{v})}{\sqrt{4{(H_{cv}^{N(F_i)})}^{2}+{(H_{c}^{N(F_i)}-H_{v}^{N(F_i)})}^2}}\right],
  \end{equation}
has the longitudinal wave vector $k_x^{N(F_i)}=[(\sqrt{\mathcal{E}_3^{N(F_i)}}-2 \gamma^2+\eta_{_{-}} \mathcal{E}_2^{N(F_i)}-\eta_{_{+}} \mathcal{E}_1 )/(\eta_{_{+}}^2-\eta_{_{-}}^2)]^{1/2}$, with ${\mathcal{E}_1}=(E_{_{-}} + k_{y}^2\nu_{+})$, $\mathcal{E}_2^{N(F_i)}= (E_{_{+}} + 2\sigma m - 2\mu_{N(F)} + k_{y}^2\nu_{-})$,  $\mathcal{E}_3^{N(F_i)}=4 \gamma^4+4 \gamma^2 (-\eta_{_{-}} \mathcal{E}_2^{N(F_i)}+ \eta_{_{+}} \mathcal{E}_1)+(\eta_{_{-}} \mathcal{E}_1-\eta_{_{+}}  \mathcal{E}_2^{N(F_i)})^2$,  $E_{_{\pm}}=E_c\pm E_v, \eta_{_{\pm}}=\eta_c\pm \eta_v$, and $\nu_{_{\pm}}=\nu_c\pm \nu_v$. The other parameters in Eqs. (\ref{psiF}), (\ref{psiN1}), and (\ref{psiN2}) are defined by
\begin{eqnarray}
&&\chi_{_{\tau}}^{N(F_i)}=\frac{\mathcal{H}_{c}^{N(F_i)}- \mathcal{H}_{v}^{N(F_i)} +\tau \sqrt{4{({\mathcal{H}}_{cv}^{N(F_i)})}^2 + {(\mathcal{H}_{c}^{N(F_i)} -{\mathcal{H}}_{v}^{N(F_i)})}^2}}{2{\mathcal{H}}_{cv}^{N(F_i)}},\nonumber\\
&&A_{\sigma,\tau}^{N(F_i)}=\sqrt{\frac{\hbar v_x^{N(F_i)}}{4\eta_c k_x^{N(F_i)}{|\chi_{_{\tau}}^{N(F_i)}|}^2+4\gamma\ Re {(\chi_{_{\tau}}^{N(F_i)})}-4\eta_v k_x^{N(F_i)}}}.\nonumber\\
\end{eqnarray}
\end{widetext}
Finally, from the obtained expressions of the transmission amplitudes $t_{\sigma\bar{\sigma}}$ and $t_{\sigma{\sigma}}$ (see the Appendix), we define the spin-current density as~\cite{Stiles}
\begin{equation}
\label{spin cuurent}
 J^{lx}_{S}(\bm{r})= \frac{\hbar}{2}\ Re \sum_{k_y } [\Psi^{\dagger} (\bm{r}) {\hat{\sigma}}_l \otimes  {\hat{\bf v}}_x \Psi(\bm{r})],
\end{equation}
where ${\hat{\bf v}}_x={\hbar^{-1}}(\partial {{\cal H}}/\partial{k_x})$, and the superscript $l$ denotes the direction of the spin Pauli matrices. From now on, since we consider only incoming and outgoing quasiparticles from the conduction bands of the F region, we thus simplify $\chi_{\tau}^{_{F_i,\sigma(\bar{\sigma})}}$ using $\chi^{_{F_i,\sigma(\bar{\sigma})}}$. The components of the spin-current density are obtained as follows:

\begin{widetext}
\begin{eqnarray}
\label{jxx}
J_S^{xx}&=&Re\{A_{{\sigma}}^{*_{F_ 2}} t_ {{\sigma} {\sigma}}^{*} (4A_{{\sigma}}^{_{F_ 2}} t_ {{\sigma} {\sigma}} Re[\chi^{_{F_ 2,{\sigma}}}](\eta_{_{-}}k_ {x}^{F_ 2,\sigma}+\gamma \cos\phi_2)- 2 A_{\bar{\sigma}}^{_{F_ 2}} t_ {{\sigma} \bar{\sigma}}e^{ix(k_ {x}^{F_ 2,\bar{\sigma}}-k_ {x}^{F_ 2,\sigma})}(\chi^{*_{F_ 2,\sigma}}+\chi^{_{F_ 2,\bar{\sigma}}})(k_ {x}^{F_ 2,\bar{\sigma}}\eta_{_{+}}-i\gamma \sin\phi_2)) \nonumber\\
&-&A_{\bar{\sigma}}^{*_{F_ 2}} t_ {{\sigma} \bar{\sigma}}^{*} (4A_{\bar{\sigma}}^{_{F_ 2}} t_ {{\sigma} \bar{\sigma}} Re[\chi^{_{F_ 2,\bar{\sigma}}}](-\eta_{_{-}}k_ {x}^{F_ 2,\bar{\sigma}}+\gamma \cos\phi_2)+ 2 A_{{\sigma}}^{_{F_ 2}} t_ {{\sigma} {\sigma}}e^{-ix(k_ {x}^{F_ 2,\bar{\sigma}}-k_ {x}^{F_ 2,\sigma})}(\chi^{*_{F_ 2,\bar{\sigma}}}+\chi^{_{F_ 2,{\sigma}}})(k_ {x}^{F_ 2,{\sigma}}\eta_{_{+}}+i\gamma \sin\phi_2))\},\nonumber\\
\end{eqnarray}

\begin{eqnarray}
\label{jyx}
J_S^{yx}&=&- Im\{A_{{\sigma}}^{*_{F_ 2}} t_ {{\sigma} {\sigma}}^{*} (4iA_{{\sigma}}^{_{F_ 2}} t_ {{\sigma} {\sigma}} Im[\chi^{_{F_ 2,{\sigma}}}](\eta_{_{+}}k_ {x}^{F_ 2,\sigma}+\gamma \cos\phi_2)+ 2 A_{\bar{\sigma}}^{_{F_ 2}} t_ {{\sigma} \bar{\sigma}}e^{ix(k_ {x}^{F_ 2,\bar{\sigma}}-k_ {x}^{F_ 2,\sigma})}(\chi^{*_{F_ 2,\sigma}}-\chi^{_{F_ 2,\bar{\sigma}}})(k_ {x}^{F_ 2,\bar{\sigma}}\eta_{_{+}}-i\gamma \nonumber\\
&\times& \sin\phi_2))+A_{\bar{\sigma}}^{*_{F_ 2}} t_ {{\sigma} \bar{\sigma}}^{*} (4i A_{\bar{\sigma}}^{_{F_ 2}} t_ {{\sigma} \bar{\sigma}} Im[\chi^{_{F_ 2,\bar{\sigma}}}](\eta_{_{-}}k_ {x}^{F_ 2,\bar{\sigma}}-\gamma \cos\phi_2)+ 2 A_{{\sigma}}^{_{F_ 2}} t_ {{\sigma} {\sigma}}e^{-ix(k_ {x}^{F_ 2,\bar{\sigma}}-k_ {x}^{F_ 2,\sigma})}(\chi^{*_{F_ 2,\bar{\sigma}}}-\chi^{_{F_ 2,{\sigma}}})(k_ {x}^{F_ 2,{\sigma}}\eta_{_{+}}\nonumber\\
&+&i\gamma \sin\phi_2))\},
\end{eqnarray}

\begin{eqnarray}
\label{jzx}
J_S^{zx}&=&Re\{A_{{\sigma}}^{*_{F_ 2}} t_ {{\sigma} {\sigma}}^{*} (2A_{{\sigma}}^{_{F_ 2}} t_ {{\sigma} {\sigma}} (|\chi^{_{F_ 2,\sigma}}|^2-1)(\eta_{_{-}}k_ {x}^{F_ 2,\sigma}+\gamma \cos\phi_2)- 2 A_{\bar{\sigma}}^{_{F_ 2}} t_ {{\sigma} \bar{\sigma}}e^{ix(k_ {x}^{F_ 2,\bar{\sigma}}-k_ {x}^{F_ 2,\sigma})}(\chi^{*_{F_ 2,\sigma}}\chi^{_{F_ 2,\bar{\sigma}}}-1)(k_ {x}^{F_ 2,\bar{\sigma}}\eta_{_{+}}-i\gamma\nonumber\\
&\times& \sin\phi_2))+A_{\bar{\sigma}}^{*_{F_ 2}} t_ {{\sigma} \bar{\sigma}}^{*} (-2A_{\bar{\sigma}}^{_{F_ 2}} t_ {{\sigma} \bar{\sigma}} (|\chi^{_{F_ 2,\bar{\sigma}}}|^2-1)(-\eta_{_{-}}k_ {x}^{F_ 2,\bar{\sigma}}+\gamma \cos\phi_2)- 2 A_{{\sigma}}^{_{F_ 2}} t_ {{\sigma} {\sigma}}e^{-ix(k_ {x}^{F_ 2,\bar{\sigma}}-k_ {x}^{F_ 2,\sigma})}(\chi^{*_{F_ 2,\bar{\sigma}}}\chi^{_{F_ 2,{\sigma}}}-1)\nonumber\\
&\times&(k_ {x}^{F_ 2,{\sigma}}\eta_{_{+}}+i\gamma\sin\phi_2))\}.
\end{eqnarray}
\end{widetext}
It is worthwhile mentioning that the anisotropic feature of the band structure appears in the aforementioned equations. Finally, having calculated the above-found expressions of $J^{lx}_{S}$, we can compute the $l$ component of the spin-transfer torque, exerted on the magnetization vector of the right F region, using the following formula,
\begin{equation}
\label{tau}
\tau_{STT}^l = {\partial J^{lx}_S}/{\partial x}.
\end{equation}
The resulting spin-transfer torque and also the charge conductance of the F/N/F junction will be discussed in the next section.

\section{Numerical results and discussions}\label{sec:level2}

In this section, we present our numerical results obtained using the numerical transmission amplitudes, and Eqs. (\ref{conductance}), (\ref{spin cuurent}), and (\ref{tau}). All of the energies $\mu_F$, $\mu_N$, and $m$ are in units of eV. We scale the length of the N region $L$, in units of the lattice constant in the $x$ (armchair) direction $a_x=4.63$ {\AA}. Note that the chemical potential is considered the same for the both F regions.

\subsection{Transmission probability}
\begin{figure}[t]
\includegraphics[width=3.5in]{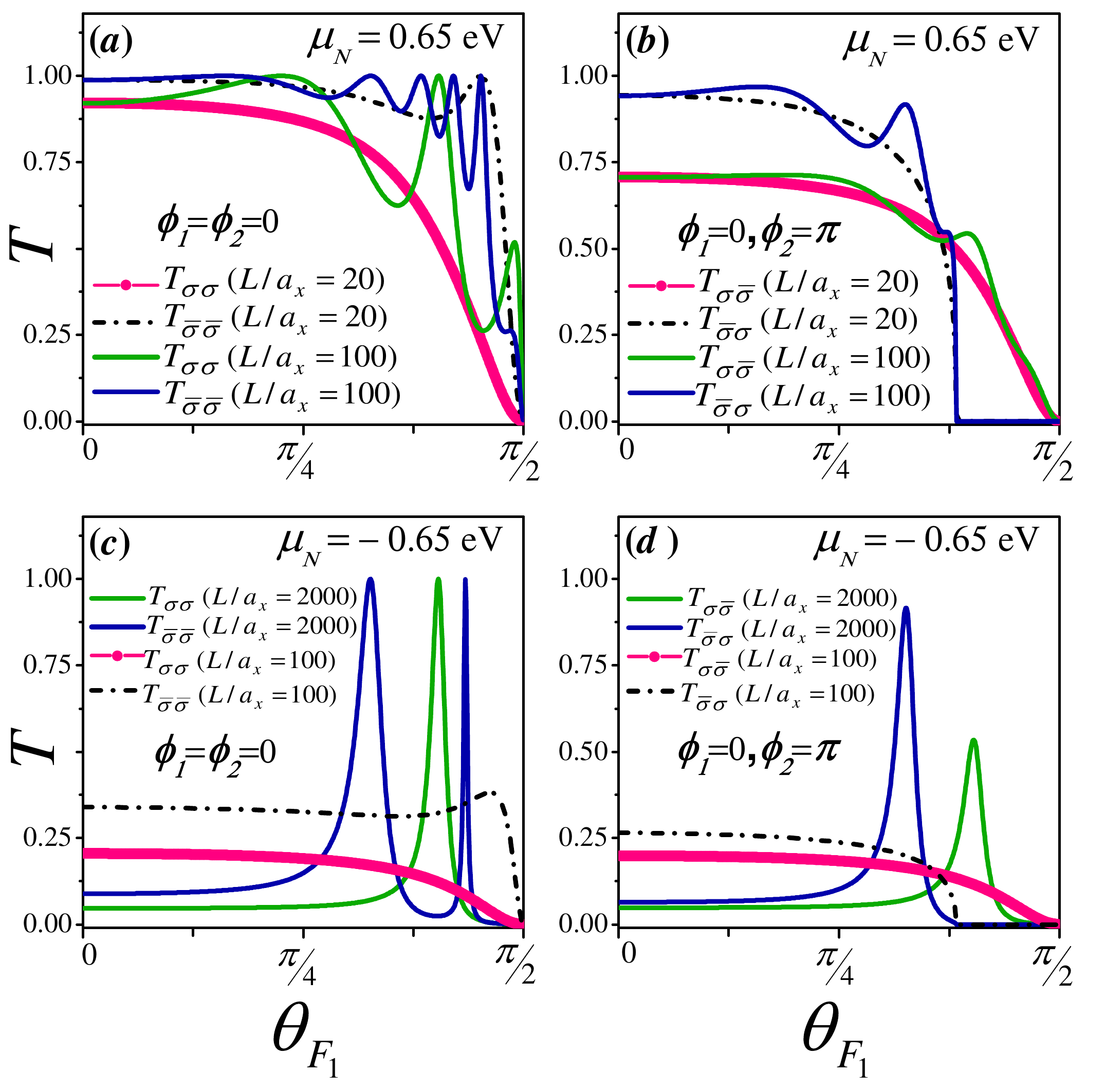}.
\caption{Top (bottom) panel: The transmission probabilities for incoming electrons from $\sigma$ and $\bar{\sigma}$ bands of the left F region in the F/N/F structure with $n$-($p$-)doped N region versus the angle of incidence $\theta_{F_1}$ for the parallel ($\phi_1=\phi_2=0$) and antiparallel ($\phi_1=0$, $\phi_2=\pi$) configuration of magnetization, when $\mu_F=0.54$ eV for both F regions and $m=0.03$ eV. The energy of the propagating electrons is equal to $\mu_F$.
\label{Fig:3}}
\end{figure}
First, we evaluate the transmission probability of the electron from the $\sigma$ ($\bar{\sigma}$) band of the left F region propagating with the energy equal to the chemical potential $\mu_F$ to the $\sigma$ ($\bar{\sigma}$) band of the right F region in the proposed F/N/F structure with different alignments of magnetization vectors. Due to the translational invariance along the $y$ direction, the $y$-component of the wave vector $k_y$ is conserved. Therefore, there will be critical angles of incidence for different transmission processes above which certain types of transmissions are forbidden.
\par
Figure \ref{Fig:3} presents the behavior of the transmission probabilities in terms of the angle of incidence $\theta_{F_1}$ for the parallel (P) and antiparallel (AP) configuration of magnetization, respectively, in the left and right panels. We set the chemical potential of the $n$-type F and $n$-($p$-)type doped N regions $\mu_F=0.54$ eV and $\mu_N=\pm 0.65$ eV, respectively, and the exchange field $m=0.03$ eV. Since the exchange field is finite, the strength of the barrier potential is different for incoming electrons from different spin subbands [see Fig. \ref{Fig:2}(a)]. In the P configuration, the transmission probability $T_{{\sigma}{\sigma}}$ ($T_{{\sigma}{\sigma}}={|t_{{\sigma}{\sigma}}|}^2$) is less than the $T_{\bar{\sigma}\bar{\sigma}}$ and monotonically decreases with $\theta_{F_1}$ for a small length of the N region ($L/a_x$), while a peak structure with unit transmission appears for $T_{\bar{\sigma}\bar{\sigma}}$ [see Fig. \ref{Fig:3}(a)]. Increasing the $L/a_x$ value leads to the appearance of a peak structure for $T_{{\sigma}{\sigma}}$ together with additional peaks with unit amplitudes for $T_{\bar{\sigma}\bar{\sigma}}$. The unity is attained in prerequisite condition when $k^N_x L=n\pi$, where $n$ is an integer value~\cite{Zheng}. In the AP configuration, the transmission probabilities of both the $\sigma$ and $\bar{\sigma}$ band electrons have decreasing behavior with $\theta_{F_1}$, and peak structures with $T_{{\sigma}\bar{\sigma}({\bar{\sigma}}{\sigma})}<1$ appear with the increase of the contact length [see Fig. \ref{Fig:3}(b)]. Note that there is a critical angle of incidence defined as $\theta_c^{\bar{\sigma}{\sigma}}=\arcsin(k'_{{\sigma}}/k'_{\bar{\sigma}})$ for the incoming electron from the $\bar{\sigma}$ band above which the corresponding wave becomes evanescent and does not contribute to any transport of charge. We find (not show) that decreasing the value of the exchange field $m$ leads to the enhancement of the transmission probabilities $T_{{\sigma}, \bar{\sigma}}$  and $T_{\bar{\sigma}, {\sigma}}$, and appearance of the peak structure with perfect transmissions. We mention that the transmission probabilities between different spin subbands ($T_{\sigma,\bar{\sigma}}$, $T_{\bar{\sigma},\sigma}$) in the P configuration and also between equal spin subbands ($T_{\sigma,\sigma}$, $T_{\bar{\sigma},\bar{\sigma}}$) in the AP configuration are equal to zero.
\par
Furthermore, the effect of a local gate voltage in the N region of the proposed structure on the transmission probabilities is shown in the bottom panel of Fig. \ref{Fig:3} for the P [Fig. \ref{Fig:3}(c)] and AP [Fig. \ref{Fig:3}(d)] configurations. It is seen that increasing the $L/a_x$ value leads to the perfect transmission, where $k^N_x L=n\pi$ is satisfied, at defined angles of incidence and attenuation of the transmission at other traveling modes, in the corresponding structure with $p$-doped N region.

\subsection{Charge conductance}
\begin{figure}[t]
\includegraphics[width=3.4in]{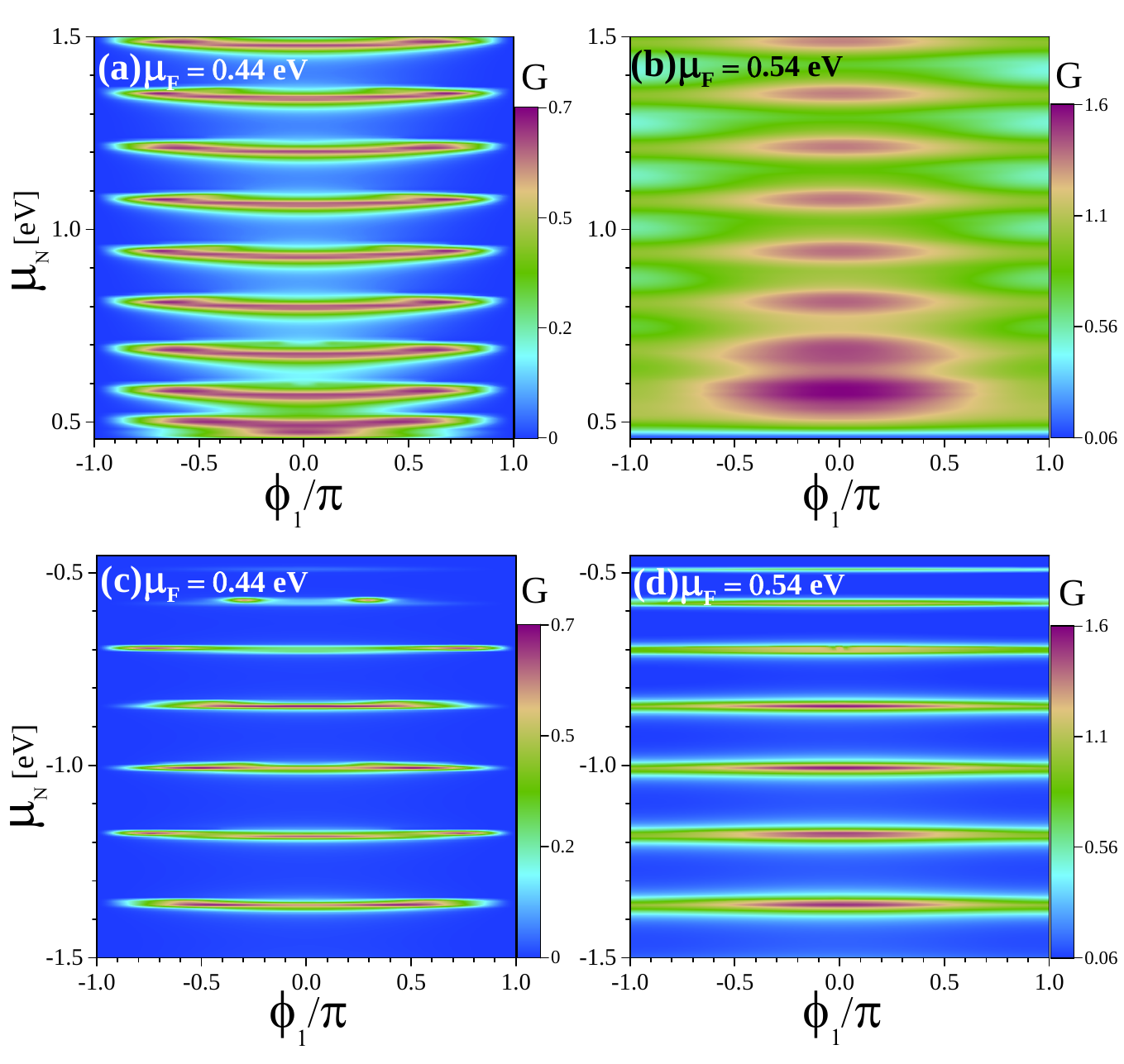}.
\caption{Top (bottom) panel: The charge conductance of the F/N/F structure with $n$-($p$-)doped N region (in units of $g_0 W/{2\pi}$) versus the magnetization direction of the left F region $\phi_1$ and the chemical potential of the N region $\mu_{N}$ for two values of the chemical potential of the F regions $\mu_F=0.44$ eV (left panel) and $\mu_F=0.54$ eV (right panel), when $m=0.03$ eV, $\phi_2=0$, and $L/a_x=20$.
\label{Fig:4}}
\end{figure}
\begin{figure}[h]
\includegraphics[width=3.4in]{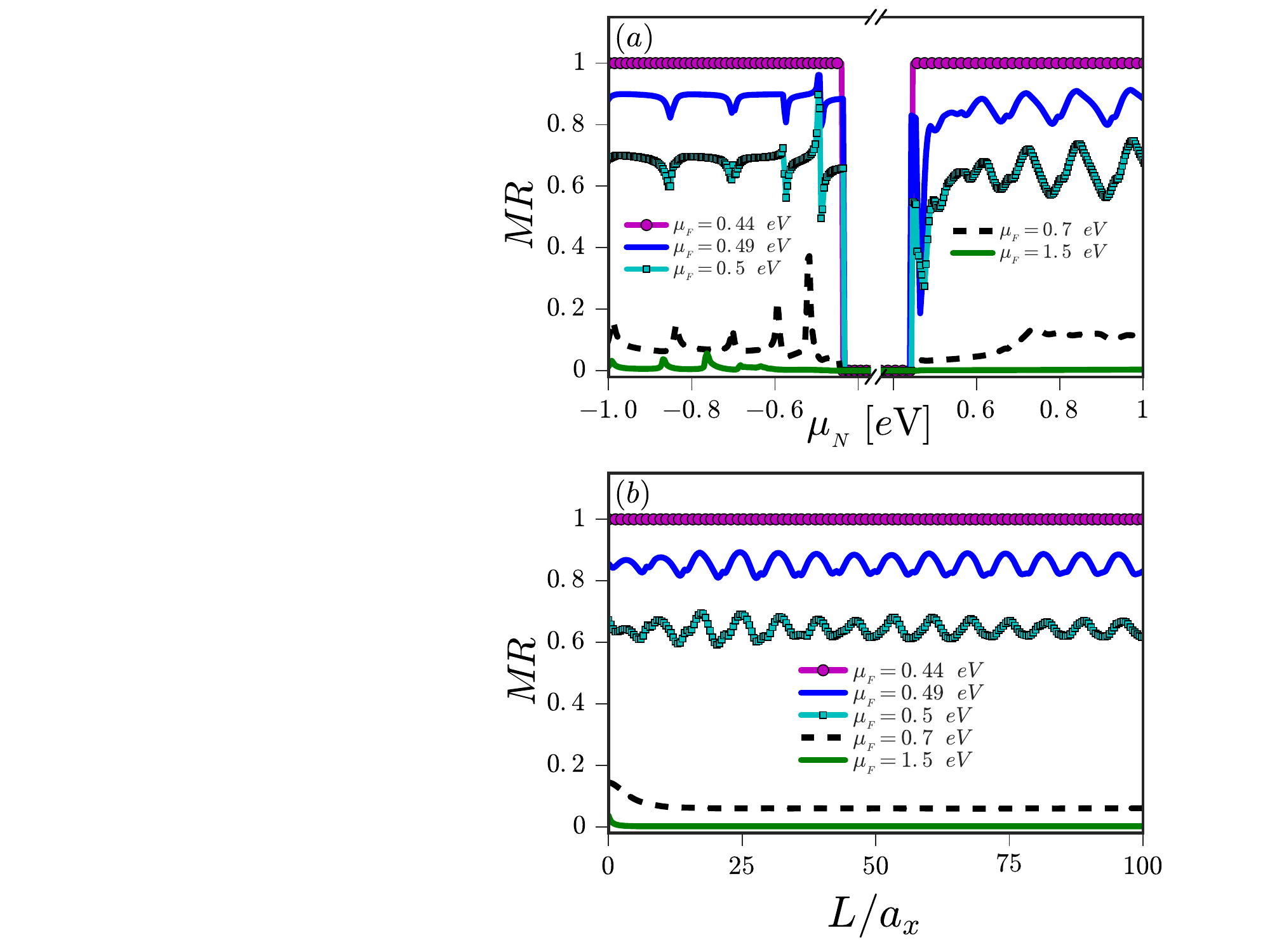}.
\caption{Magnetoresistance of the spin valve structure versus the chemical potential $\mu_N$ for $m=0.03$ eV, when $L/a_x=20$ (a) and the length of the N region $L/a_x$, when $\mu_N=0.65$ eV (b) for different values of $\mu_F$.
\label{Fig:5}}
\end{figure}
Now, we proceed to investigate the charge conductance of the proposed F/N/F structure at zero temperature. We consider two cases for the chemical potential of the F region: $E^{\bar{\sigma}, \tau}_{BE}\leq\mu_F<E^{{\sigma}, \tau}_{BE}$ and $\mu_F\geq E^{{\sigma}, \tau}_{BE}$ (see Fig. \ref{Fig:2}). The energies $E^{\bar{\sigma}, \tau}_{BE}$ and $E^{{\sigma}, \tau}_{BE}$ define the conduction-band edges for the two spin-subbands of the F region. The incoming quasiparticles from the left F region with the chemical potential in the range $E^{\bar{\sigma}, \tau}_{BE}\leq\mu_F<E^{{\sigma}, \tau}_{BE}$ are completely dominated by the electrons from the $({\bar{\sigma}, \tau})$ band with $\bar{\sigma}=-1$ and $\tau=1$ while for the case of $\mu_F\geq E^{{\sigma}, \tau}_{BE}$, the incoming electrons are from both $({\sigma, \tau})$ and $({\bar{\sigma}, \tau})$ bands with ${\sigma}=-\bar{\sigma}=1$ and $\tau=1$.
\par
The behavior of the charge conductance of the $n$-doped F/N/F structure (in units of $g_0 W/{2\pi}$) versus the chemical potential of the N region, $\mu_N$, and the magnetization direction of the left F region, $\phi_1$, are demonstrated in Figs. \ref{Fig:4}(a) and \ref{Fig:4}(b) for $\mu_F = 0.44$ eV ($\mu_F<E^{{\sigma}, \tau}_{BE}$) and $\mu_F = 0.54$ eV ($\mu_F\geq E^{{\sigma}, \tau}_{BE}$), when the magnetization direction of the right F region, $\bm{m}_2$, is fixed to $\phi_2=0$, and $L/a_x=20$. As seen, the charge conductance is an even function of $\phi_1$ and oscillates by varying $\mu_N$ with the gate voltage. The oscillations can be understood in the context of the Fabry-P$\acute{e}$rot oscillations~\cite{Born}. Interestingly, the charge conductance can be very small ($\sim$ zero) away from the AP configuration of magnetization ($\phi_1=\pm \pi$) depending on the value of $\mu_N$, when $\mu_F = 0.44$ eV [see Fig. \ref{Fig:4}(a)], while in the case of $\mu_F=0.54$ eV, it is seen from Fig. \ref{Fig:4}(b) that the charge conductance is finite even at the AP configuration. The corresponding results for the case of the $p$-doped N region are plotted in the bottom panel of Fig. \ref{Fig:4} [Figs. \ref{Fig:4}(c) and \ref{Fig:4}(d)]. For $\mu_F=0.44$ eV, the charge conductance is finite for small ranges of $\mu_N$. For these values of $\mu_N$, interestingly it is seen that the charge conductance can be reduced at the P configuration and enhanced near the AP configuration. Importantly, we find that the charge conductance of the case $\mu_F= 0.54$ eV is finite for small ranges of $\mu_N$ in contrast to the corresponding n-doped structure, where the charge conductance can not be suppressed for AP configuration. These results reflect the anisotropic behavior of the quasiparticles in the conduction and valance bands of the monolayer phosphorene.
\par
With the above-found behavior of the charge conductance, we evaluate the magnetoresistance of the proposed spin valve structure defined as the relative difference of the charge conductance in the two P and AP configurations, $MR=(G_P-G_{AP})/G_P$. The behavior of the resulting MR in terms of $\mu_N$ and the length of the N region are  presented respectively in Figs. \ref{Fig:5}(a) and \ref{Fig:5}(b), for different values of $\mu_F$. We observe that the spin valve effect can be perfect (with $MR=1$) for the case of $\mu_F=0.44$ eV ($\mu_F<E^{{\sigma}, \tau}_{BE}$). This perfect switching is robust with respect to an increase of the chemical potential of the N region. Increasing $\mu_F$ leads to the oscillatory behavior of the MR with respect to $\mu_N$, such that the period and amplitude of the oscillations differ for the two cases of n-type and p-type doped N regions. The amplitude of the MR decreases with increasing $\mu_F$ and tends to zero for the highly doped F region. Note that the existence of a band gap in the dispersion of the monolayer phosphorene causes a gap in MR for $E^{{\sigma}, \bar{\tau}}_{BE}<\mu_N<E^{\bar{\sigma}, \tau}_{BE}$ ($E^{{\sigma}, \bar{\tau}}_{BE}$ defines the energy of the valance-band edge for the spin ${\sigma}$ band). Moreover, it is shown in Fig. \ref{Fig:5}(b) that the perfect spin valve effect is preserved for large lengths of the N region, when $\mu_F=0.44$ eV. This is similar to the behavior of the pseudo-magnetoresistance in the monolayer-graphene-based pseudospin valve structure~\cite{Majidi11,Majidi13}. For $\mu_F>E^{{\sigma}, \tau}_{BE}$, MR shows an oscillatory behavior with respect to the $L/a_x$ value with the amplitude which decreases with $\mu_F$ and goes to zero for high values of $\mu_F$. The oscillatory behavior of the MR with respect to the chemical potential $\mu_N$ and the length $L$ can be explained in the context of the Fabry-P$\acute{e}$rot oscillations~\cite{Born}, too.

\subsection{Spin-transfer torque}

\begin{figure}[]
\includegraphics[width=3.4in]{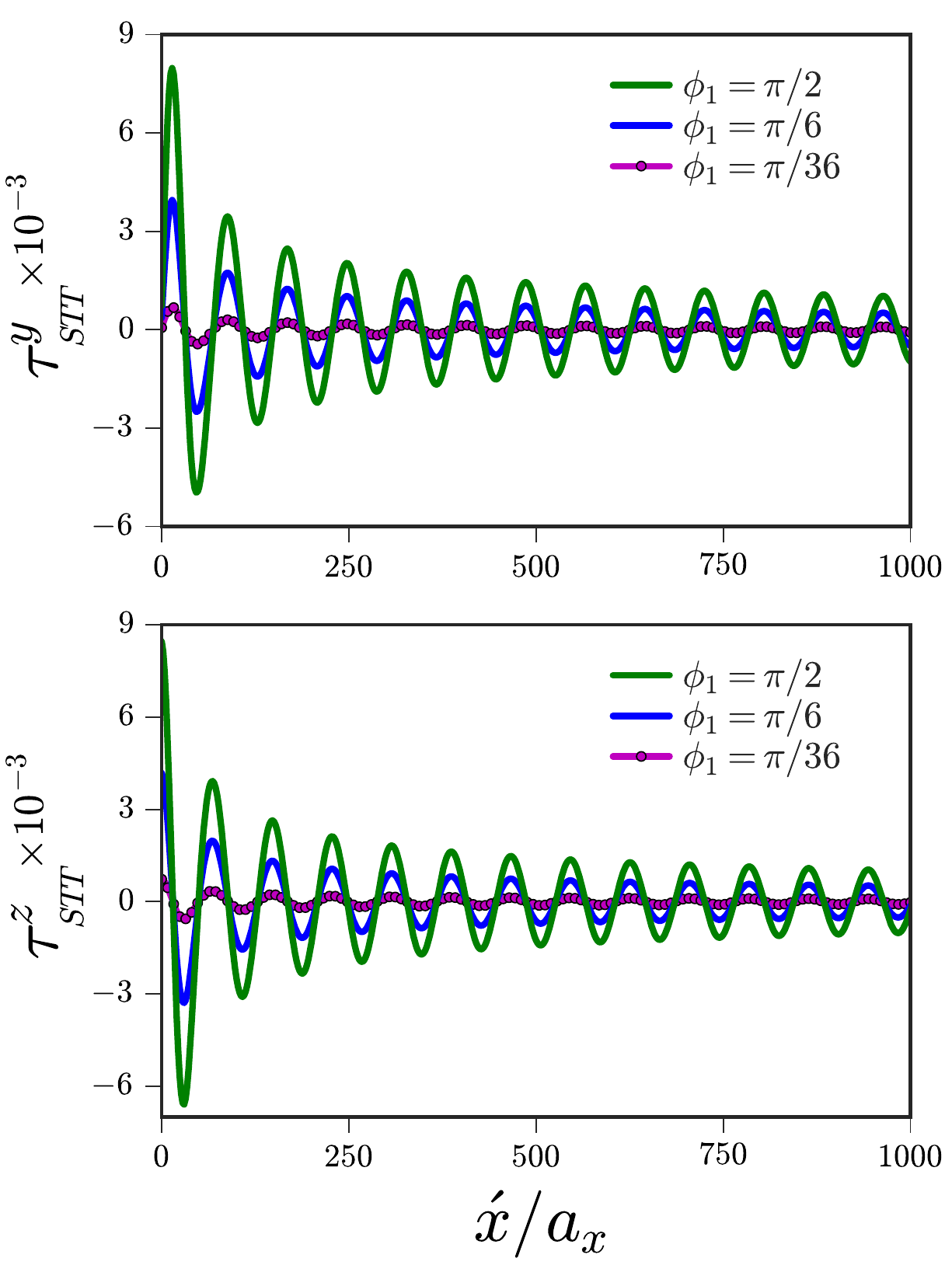}.
\caption{The spin-transfer torques (a) ${\tau}_{STT}^y$ and (b) ${\tau}_{STT}^z$ (in units of ${W}/{4\pi}$) as a function of the distance from the N/F interface for three different values of $\phi_1$, when $\mu_{F}=0.44$ eV, $\mu_{N}=0.65$ eV, $m=0.03$ eV, $\phi_2=0$, and $L/a_x=20$.
\label{Fig:6}}
\end{figure}
\begin{figure}[]
\includegraphics[width=3.4in]{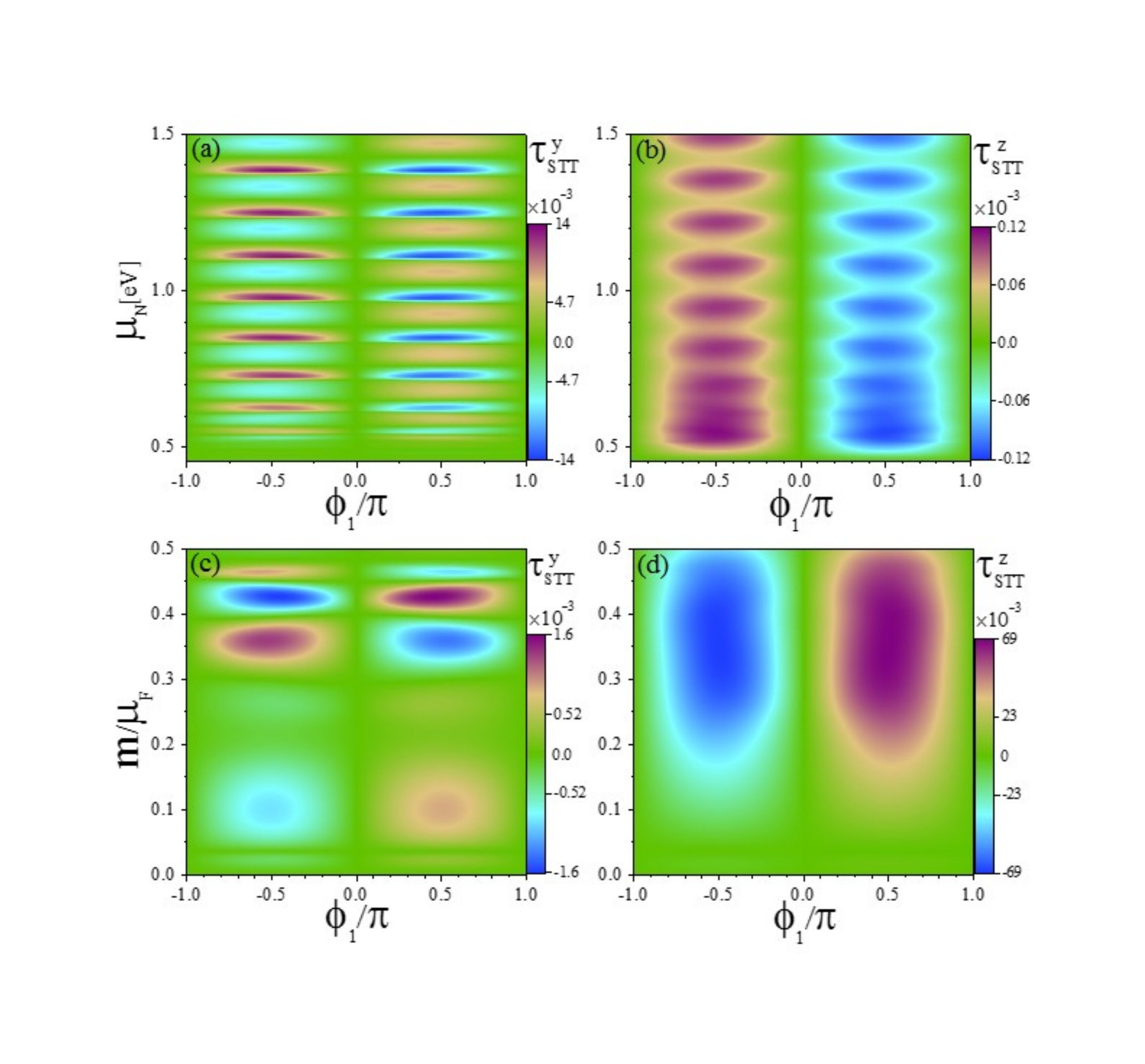}.
\caption{Top panel: The spin-transfer torques (a) ${\tau}_{STT}^y$ and (b) ${\tau}_{STT}^z$ (in units of ${W}/{4\pi}$) versus the orientation of the magnetization vector $\bm{m}_1$, $\phi_1$, and the chemical potential of the N region $\mu_N$, when $m=0.03$ eV. Bottom panel: (c) ${\tau}_{STT}^y$ and (d) ${\tau}_{STT}^z$ versus $\phi_1$ and the magnitude of the exchange field $m/\mu_F$ (in units of the chemical potential of the F region), when $\mu_N= 0.65$ eV. The magnetization vector $\bm{m}_2$ is fixed to $x$ axis ($\phi_2$=0), $\mu_F=0.44$ eV, and $L/a_x=20$.
\label{Fig:7}}
\end{figure}
Finally, we evaluate the behavior of the spin-transfer torque exerted on the magnetic order parameter of the right F region in the proposed F/N/F structure. Three-layer structures consisting of two magnetic layers separated by a conducting nonmagnetic spacer layer are well-known systems that exploit spin-transfer torque effects. When applying a charge current perpendicular to the layers, one of the magnetic layers, referred to as the pinned layer, acts as a spin polarizer. When the spin-polarized electrons reach the second layer, referred to as the free layer, they accumulate at the interface and thereby exert a torque onto the free layer.

Figure \ref{Fig:6} presents the dependence of the spin-transfer torques ${\tau}_{STT}^y$ and ${\tau}_{STT}^z$ (in units of $W/{4\pi}$) on the distance $\acute{x}$ ($\acute{x}=x-L$) from the N/F interface for different values of $\phi_1$. We fix $\phi_2=0$, so that the component of the torque parallel to the magnetization of the right F region, ${\tau}_{STT}^x$, is zero. As seen, the total magnetic torque is not absorbed at the N/F interface and penetrates into the right F region. The torque components exhibit especially damped oscillatory behavior and their amplitudes diminish with decreasing $\phi_1$. It is seen from Eqs. (\ref{jxx}), (\ref{jyx}), (\ref{jzx}), and (\ref{tau}) that the torque components possess the terms proportional to $e^{i(k_ {x}^{F_ 2,\bar{\sigma}}-k_ {x}^{F_ 2,\sigma})x}$ which cause the conventional rapid oscillations on a length scale $2\pi(k_ {x}^{F_ 2,\bar{\sigma}}-k_ {x}^{F_ 2,\sigma})\sim 1/m$. These oscillations originate from the interference between electrons with opposite spin-indices. Moreover, we find that the amplitudes of the torque components decay with ${\acute{x}}^{-1}$. The penetration of the spin transfer torque into the right F region in the phosphorene-based F/N/F structure is similar to that in the ferromagnetic superconductor-~\cite{FS} and graphene-based ~\cite{fnfgraphene1} spin valve structures, where the vast majority of the torque is absorbed in the interface region.

The top panel of Fig. \ref{Fig:7} shows the behavior of the torque components absorbed at the N/F interface ($\acute{x}=0$) in terms of $\mu_N$ and $\phi_1$, when the magnetization orientation of the right F region is fixed to $\phi_2=0$. The spin-transfer torque is seen to be an odd function of $\phi_1$ and oscillates with $\mu_N$. The sign and magnitude of ${\tau}_{STT}^y$ can be modulated by tuning $\mu_N$ with the gate voltage, while the sign of ${\tau}_{STT}^z$ can-not be changed. Also, the magnitude of ${\tau}_{STT}^y$ is two orders of magnitudes larger than ${\tau}_{STT}^z$. Interestingly, changing the exchange field $m/\mu_F$ value leads to the sign change of the torque ${\tau}_{STT}^y$ in addition to its magnitude changes [see Fig. \ref{Fig:7}(c)]. However, the magnitude of the torque component ${\tau}_{STT}^z$ increases with increasing the $m/\mu_F$ value, and can be two orders of magnitude larger than that of the ${\tau}_{STT}^y$. A large discrepancy between ${\tau}_{STT}^z$ and ${\tau}_{STT}^y$ originates from the anisotropic band structure of phosphorene. Note that in the graphene F/N/F structure, ${\tau}_{STT}^z$ is qualitatively similar to ${\tau}_{STT}^y$~\cite{fnfgraphene1}.

\section{Conclusion}\label{sec:level3}
In conclusion, we have investigated the magnetotransport characteristics of a monolayer phosphorene ferromagnetic/normal/ferromagnetic (F/N/F) junction with noncollinear magnetization. We have found that the transmission probabilities of the incoming electrons from the spin $\sigma$ and $\bar{\sigma}$ bands of the F region have different behavior by increasing the angle of incidence to the F/N/F structure with parallel (P) alignment of magnetization vectors, when the length of the N region $L$ is small. The transmission probability has a monotonically decreasing behavior with the incidence angle for an incoming ${\sigma}$-band electron, while a peak structure appears for an incident $\bar{\sigma}$-band electron. Increasing the contact length $L$ leads to the appearance of a peak structure with unit transmission for the incoming ${\sigma}$-band electron and the enhancement of the peak number in the transmission probability of the incoming $\bar{\sigma}$-band electron, while the transmission probabilities $T_{{\sigma}\bar{\sigma}}$ and $T_{{\bar{\sigma}}{\sigma}}$, of both the incoming $\sigma$- and $\bar{\sigma}$-band electrons to the proposed structure with anti-parallel (AP) alignment of magnetization have decreasing behavior with respect to the angle of incidence for small contact lengths and peak structures with the amplitudes $T_{{\sigma}\bar{\sigma}({\bar{\sigma}}{\sigma})}<1$ for larger lengths $L$. Also, applying a local gate voltage to the N region with large length $L$ amplifies the transmission probability at defined angles of incidence and attenuates it at other traveling modes in the corresponding F/N/F structure with the $p$-doped N region.
\par
Moreover, we have shown that the charge conductance of the proposed structure can be changed significantly by tuning the chemical potential of the N region $\mu_N$, and the relative orientation of the magnetization directions in the two F regions. Interestingly, the charge conductance of the n-type doped F/N/F structure can be almost zero away from the AP configuration of magnetization, for determining ranges of the chemical potential of the F region $\mu_F$. More importantly, it attains a minimum at the P configuration and a maximum near the AP configuration in the corresponding structure with $p$-type doped N region. We have further demonstrated that the proposed structure exhibits a relative difference of the electrical resistance in the P and AP alignments of magnetization, which can be tuned to unity for defined values of $\mu_F$. This perfect spin valve effect is more robust with respect to an increase of $\mu_N$ and is preserved even for large lengths of the N region. Furthermore, we have evaluated the equilibrium spin-transfer torque, which can be used to derive domain-wall motion in ferromagnetic materials, acting on the magnetization of the right F region. We have found that the total torque is not absorbed in the interface region, and both the magnitude and sign of the torque components can be controlled by tuning $\mu_N$ and the exchange field of the F region.

\section{Acknowledgement}

This work is partially supported by Iran Science Elites Federation.

\section{Appendix: spin-dependent transmission amplitudes}

As said before, we calculate the transmission amplitudes $t_{\sigma{\sigma}}$ and $t_{\sigma\bar{\sigma}}$ by matching the wave functions of the three regions of the left F region, the N region, and the right F region at the two interfaces, namely $x=0$ and $x=L$. Those amplitudes read as
\begin{equation}
t_{\sigma\sigma(\bar{\sigma})}=\frac{f_{\sigma\sigma(\bar{\sigma})}}{{f'}_{\sigma\sigma(\bar{\sigma})}},
\end{equation}
with
\begin{widetext}
\begin{equation}
\label{f1}
f_{\sigma\sigma} =-2A_{\sigma}^{F_1}\chi^{_{F_1,\sigma}} {\chi_\tau^{_N}}e^{i(k_x^{N}L+\phi_{2})}(e^{i \phi_{1}}+e^{i \phi_{2}})[- (\chi^{_{F_1,\bar{\sigma}}}+{\chi_\tau^{_N}})(\chi^{_{F_2,\bar{\sigma}}}+{\chi_\tau^{_N}})+({\chi_\tau^{_N}}-\chi^{_{F_1,\bar{\sigma}}})({\chi_\tau^{_N}}-
\chi^{_{F_2,\bar{\sigma}}})e^{2ik_x^{N}L}],\\
\end{equation}
\end{widetext}
\begin{widetext}
\begin{eqnarray}
\label{f'1}
f'_{\sigma\sigma} &=& A_{\sigma}^{F_2} e^{ik_ {x}^{F_ 2,\sigma}L}[2(\chi^{_{F_1,\bar{\sigma}}}\chi^{_{F_1,\sigma}} - {\chi_\tau^{_N}}^2)({\chi_\tau^{_N}}^2 - \chi^{_{F_2,\bar{\sigma}}}\chi^{_{F_2,\sigma}})e^{i(2k_x^{N}L + \phi_{1} + \phi_{2})} -{\chi_\tau^{_N}}^2(\chi^{_{F_1,\bar{\sigma}}}-\chi^{_{F_1,\sigma}} )(\chi^{_{F_2,\bar{\sigma}}}-\chi^{_{F_2,\sigma}})\nonumber\\
&\times& e^{2i(k_x^{N}L + \phi_{1})}-{\chi_\tau^{_N}}^2(\chi^{_{F_1,\bar{\sigma}}}-\chi^{_{F_1,\sigma}} )(\chi^{_{F_2,\bar{\sigma}}}-\chi^{_{F_2,\sigma}}) e^{2i(k_x^{N}L + \phi_{2})}+(\chi^{_{F_1,\bar{\sigma}}}-{\chi_\tau^{_N}})(\chi^{_{F_1,\sigma}}-{\chi_\tau^{_N}})(\chi^{_{F_2,\bar{\sigma}}}-{\chi_\tau^{_N}})\nonumber\\
&\times&(\chi^{_{F_2,\sigma}}-{\chi_\tau^{_N}}) e^{i(4k_x^{N}L + \phi_{1}+\phi_{2})}+(\chi^{_{F_1,\bar{\sigma}}}+{\chi_\tau^{_N}})(\chi^{_{F_1,\sigma}}+{\chi_\tau^{_N}})(\chi^{_{F_2,\bar{\sigma}}}+{\chi_\tau^{_N}})(\chi^{_{F_2,\sigma}}+{\chi_\tau^{_N}})e^{i( \phi_{1} + \phi_{2})}],
\end{eqnarray}
\end{widetext}
\begin{widetext}
\begin{equation}
\label{f2}
f_{\sigma\bar{\sigma}}
=2A_{\sigma}^{F_1}\chi^{_{F_1,\sigma}} {\chi_\tau^{_N}}e^{i(k_x^{N}L+\phi_{2})}(\sigma e^{i \phi_{1}}-e^{i \phi_{2}})[- (\chi^{_{F_1,\bar{\sigma}}}+{\chi_\tau^{_N}})(\chi^{_{F_2,{\bar{\sigma}}}}+{\chi_\tau^{_N}})+({\chi_\tau^{_N}}-\chi^{_{F_1,\bar{\sigma}}})({\chi_\tau^{_N}}-
\chi^{_{F_2,{\bar{\sigma}}}})e^{2ik_x^{N}L}],\\
\end{equation}
\end{widetext}
\begin{widetext}
\begin{eqnarray}
\label{f'2}
f'_{\sigma\bar{\sigma}} &=& A_{\sigma}^{F_2} e^{ik_ {x}^{F_ 2,\sigma}L}[2(\chi^{_{F_1,\bar{\sigma}}}\chi^{_{F_1,\sigma}} - {\chi_\tau^{_N}}^2)({\chi_\tau^{_N}}^2 - \chi^{_{F_2,\bar{\sigma}}}\chi^{_{F_2,\sigma}})\sigma e^{i(2k_x^{N}L + \phi_{1} + \phi_{2})} -{\chi_\tau^{_N}}^2(\chi^{_{F_1,\bar{\sigma}}}-\chi^{_{F_1,\sigma}} )(\chi^{_{F_2,\bar{\sigma}}}-\chi^{_{F_2,\sigma}})\nonumber\\
&\times& \sigma e^{2i(k_x^{N}L + \phi_{1})}-{\chi_\tau^{_N}}^2(\chi^{_{F_1,\bar{\sigma}}}-\chi^{_{F_1,\sigma}} )(\chi^{_{F_2,\bar{\sigma}}}-\chi^{_{F_2,\sigma}}) e^{2i(k_x^{N}L + \phi_{2})}+(\chi^{_{F_1,\bar{\sigma}}}-{\chi_\tau^{_N}})(\chi^{_{F_1,\sigma}}-{\chi_\tau^{_N}})(\chi^{_{F_2,\bar{\sigma}}}-{\chi_\tau^{_N}})\nonumber\\
&\times&(\chi^{_{F_2,\sigma}}-{\chi_\tau^{_N}}) \sigma e^{i(4k_x^{N}L + \phi_{1}+\phi_{2})}+(\chi^{_{F_1,\bar{\sigma}}}+{\chi_\tau^{_N}})(\chi^{_{F_1,\sigma}}+{\chi_\tau^{_N}})(\chi^{_{F_2,\bar{\sigma}}}+{\chi_\tau^{_N}})(\chi^{_{F_2,\sigma}}+{\chi_\tau^{_N}})\sigma e^{i( \phi_{1} + \phi_{2})}].
\end{eqnarray}
\end{widetext}
Finally, the spin-dependent transmission probabilities are obtained from the above relations as follows:
\begin{equation}
\label{t1}
T_{{\sigma}{\sigma}(\bar{\sigma})}=\left|t_{{\sigma}{\sigma}(\bar{\sigma})}\right|^2.
\end{equation}
We mention that the transmission probabilities of the incoming spin $\bar{\sigma}$-band electron, $T_{\bar{\sigma}\bar{\sigma}({\sigma})}=\left|t_{\bar{\sigma}\bar{\sigma}({\sigma})}\right|^2$, can be obtained from the above formulas by replacing $\sigma$ with $\bar{\sigma}$, and vice versa.


\begin{thebibliography}{99}

\bibitem{Li14}
L. Li, Y. Yu, G. J. Ye, Q. Ge, X. Ou, H. Wu, D. Feng, X. H. Chen and Y. Zhang, Nat. Nanotechnol. {\bf 9}, 372 (2014).

\bibitem{Liu14}
H. Liu, A. T. Neal, Z. Zhu, Z. Luo, X. Xu, D. Tomanek, and P. D. Ye, ACS Nano {\bf 8}, 4033 (2014).

\bibitem{andre}
A. Castellanos-Gomez, Nat. Photonics {\bf 10} 202 (2016);  F. Xia, H. Wang, D. Xiao, M. Dubey, and A. Ramasubramaniam, {\it ibid}, {\bf 8}, 899 (2014).

\bibitem{Liu15}
H. Liu, Y. Du, Y. Deng, P. D. Ye, Chem. Soc. Rev. {\bf 44}, 2732 (2015).

\bibitem{Das14}
S. Das, M. Demarteau, A. Roelofs, ACS Nano {\bf 8}, 11730 (2014).

\bibitem{Kamalakar15}
M. V. Kamalakar, B. N. Madhushankar, A. Dankert, S. P. Dash, Small {\bf 11}, 2209 (2015).

\bibitem{Lu14}
W. Lu, H. Nan, J. Hong, Y. Chen, C. Zhu, Z. Liang, X. Ma, Z. Ni, C. Jin, and Z. Zhang, Nano Res. {\bf 7}, 853 (2014).

\bibitem{Fei14}
R. Fei and L. Yang, Nano Lett. {\bf 14}, 2884 (2014).

\bibitem{Wang1}
X. Wang, A. M. Jones, K. L. Seyler, V. Tran, Y. Jia, H. Zhao, H. Wang, L. Yang, X. Xu, and F. Xia, Nat. Nanotechnol. {\bf 10}, 517 (2015).

\bibitem{Du10}
Y. Du, C. Ouyang, S. Shi, and M. Lei, J. Appl. Phys. {\bf 107}, 093718 (2010).

\bibitem{Tran14}
V. Tran, R. Soklaski, Y. Liang, and L. Yang, Phys. Rev. B {\bf 89}, 235319 (2014).

\bibitem{cakir14}
D. \c{C}akir, H. Sahin, and F. M. Peeters, Phys. Rev. B {\bf 90}, 205421 (2014).

\bibitem{Liang14}
L. Liang, J. Wang, W. Lin, B. G. Sumpter, V. Meunier, and M. Pan, Nano Lett. {\bf 14}, 6400�6406 (2014).

\bibitem{Rodin14}
A. S. Rodin, A. Carvalho, and A. H. Castro Neto, Phys. Rev. Lett. {\bf 112}, 176801 (2014).

\bibitem{Zhu14}
Z. Zhu, Ch. Li, W. Yu, D. Chang, Q. Sun, and Y. Jia, Appl. Phys. Lett. {\bf 105}, 113105 (2014).

\bibitem{Du15}
Y. Du, H. Liu, B. Xu, L. Sheng, J. Yin, C.-G. Duan, and X. Wan, Sci. Rep. {\bf 5}, 8921 (2015).

\bibitem{Yang16}
G. Yang, S. Xu, W. Zhang, T. Ma, and C. Wu, Phys. Rev. B {\bf 94}, 075106 (2016).

\bibitem{Farooq15}
M. U. Farooq, A. Hashmi, and J. Hong, ACS Appl. Mater. Interfaces {\bf 7}, 14423 (2015).

\bibitem{Farooq16}
M. U. Farooq, A. Hashmi, and J. Hong, Sci. Rep. {\bf 6}, 26300 (2016).

\bibitem{Hu15}
T. Hu and J. Hong, J. Phys. Chem. C {\bf 119}, 8199 (2015).

\bibitem{Sui15}
X. Sui, C. Si, B. Shao, X. Zou, J. Wu, B.-L. Gu, and W. Duan, J. Phys. Chem. C {\bf 119}, 10059 (2015).

\bibitem{Seixas15}
L. Seixas, A. Carvalho, and A. H. Castro Neto, Phys. Rev. B {\bf 91}, 155138 (2015).

\bibitem{khan15}
I. Khan and J. Hong, New J. Phys. {\bf 17}, 023056 (2015).

\bibitem{fabian}
I. Zuti\'{c}, J. Fabian, and S. Das Sarma, Rev. Mod. Phys. {\bf 76}, 323 (2004).

\bibitem{Berger}
L. Berger, Phys. Rev. B {\bf 54}, 9353 (1996).

\bibitem{Slonczewski}
J. C. Slonczewski, J. Magn. Magn. Mater. {\bf 159}, L1 (1996).

\bibitem{Brataas}
A. Brataas, A. D. Kent, and H. Ohno, Nat. Mater. {\bf 11}, 372 (2012).

\bibitem{Parkin}
S. S. P. Parkin et al., J. Appl. Phys. {\bf 85}, 5828 (1999); L. Thomas, G. Jan, J. Zhu, H. Liu, Y.-J. Lee, S. Le, R.-Y. Tong, K. Pi, Y.-J.
Wang, D. Shen, R. He, J. Haq, J. Teng, V. Lam, K. Huang, T. Zhong, T. Torng,
and P.-K. Wang, {\it ibid}, {\bf 115}, 172615 (2014); K. Ando, S. Fujita, J. Ito, S. Yuasa, Y. Suzuki, Y. Nakatani, T. Miyazaki, and H. Yoda, {\it ibid}, {\bf 115}, 172607 (2014).

\bibitem{Zhu08}
J.-G. Zhu, X. Zhu, and Y. Tang, IEEE Trans. Magn. {\bf 44}, 125 (2008); J.-G. Zhu, and Y. Wang, {\it ibid}. {\bf 46}, 751 (2010).

\bibitem{Tulapurkar}
A. A. Tulapurkar, Y. Suzuki, A. Fukushima, H. Kubota, H. Maehara, K. Tsunekawa, D. D. Djayaprawira, N. Watanabe, and S. Yuasa, Nature (London) {\bf 438}, 339 (2005).

\bibitem{Fert}
A. Fert, Rev. Mod. Phys. {\bf 80}, 1517 (2008).

\bibitem{Bauer}
G. E. W. Bauer, E. Saitoh, and B. J. van Wees, Nat. Mater. {\bf 11}, 391 (2012).

\bibitem{linder15}
J. Linder, and J. W. A. Robinson, Nat. Phys. {\bf 11}, 307 (2015).

\bibitem{exp1}
M. N. Baibich, J. M. Broto, A. Fert, F. Nguyen Van Dau, F. Petroff, P. Etienne, G. Creuzet, A. Friederich, and J. Chazelas,
Phys. Rev. Lett. {\bf 61}, 2472 (1988).

\bibitem{exp2}
G. Binasch, P. Gr\"{u}nberg, F. Saurenbach, and W. Zinn, Phys. Rev. B {\bf 39}, 4828 (1989).

\bibitem{Glazov}
 M. M. Glazov, P. S. Alekseev, M. A. Odnoblyudov, V. M. Chistyakov, S. A. Tarasenko, and I. N. Yassievich,  Phys. Rev. B {\bf 71}, 155313 (2005).

\bibitem{Zheng}
Z. Zheng, Y. Qi, D. Y. Xing, and J. Dong, Phys. Rev. B {\bf 59}, 14505 (1999).

\bibitem{Ertler}
C. Ertler, and J. Fabian, Appl. Phys. Lett. {\bf 89}, 242101 (2006).

\bibitem{Chatterji}
N. Chatterji, A. A. Tulapurkar, and B. Muralidharan, Appl. Phys. Lett. {\bf 105}, 232410 (2014).

\bibitem{Nguyen}
H. S. Nguyen, D. Vishnevsky, C. Sturm, D. Tanese, D. Solnyshkov, E. Galopin, A. Lema\^{i}tre, I. Sagnes, A. Amo, G. Malpuech, and J. Bloch, Phys. Rev. Lett. {\bf 110}, 236601 (2013).

\bibitem{Yuasa}
S. Yuasa, T. Nagahama, and Y. Suzuki, Science {\bf 297}, 234 (2002).

\bibitem{Petukhov}
 A G. Petukhov, A. N. Chantis and  D. O. Demchenko, Phys. Rev. Lett. {\bf 89}, 107205 (2002).

\bibitem{datta}
S. Datta, {\it Electronic transport in mesoscopic systems} (Cambridge University Press, Cambridge, England, 1995); Y. V. Nazarov and Y. Blanter, {\it Quantum Transport: Introduction to Nanoscience} (Cambridge University Press, Cambridge, England, 2009)

\bibitem{fnfgraphene1}
T. Yokoyama and J. Linder, Phys. Rev. B {\bf 83}, 081418 (2011).

\bibitem{fnfgraphene2}
J. Zou, G. Jin, and Y. Ma, J. Phys. Cond. Matt. {\bf 21}, 126001 (2009).

\bibitem{fnfsilicene}
R. Saxena, A. Saha, and S. Rao, Phy. Rev. B {\bf 92}, 245412 (2015).

\bibitem{Tombros07}
N. Tombros, C. Jozsa, M. Popinciuc, H. T. Jonkman, and B. J. van Wees, Nature (London) {\bf 448}, 571 (2007).

\bibitem{Haugen}
H. Haugen, D. Huertas-Hernando, and A. Brataas, Phys. Rev. B {\bf 77}, 115406 (2008).

\bibitem{Swartz12}
A. G. Swartz, P. M. Odenthal, Y. Hao, R. S. Ruoff, and R. K. Kawakami, ACS Nano {\bf 6}, 10063 (2012).

\bibitem{Wang15}
Z. Wang, C. Tang, R. Sachs, Y. Barlas, and J. Shi, Phys. Rev. Lett. {\bf 114}, 016603 (2015).

\bibitem{Dugaev06}
V. K. Dugaev, V. I. Litvinov, and J. Barnas, Phys. Rev. B {\bf 74}, 224438 (2006).

\bibitem{Uchoa08}
B. Uchoa, V. N. Kotov, N. M. R. Peres, and A. H. Castro Neto, Phys. Rev. Lett. {\bf 101}, 026805 (2008).

\bibitem{Yazyev10}
O. V. Yazyev, Rep. Prog. Phys. {\bf 73}, 056501 (2010).

\bibitem{Zare}
M. Zare, B. Z. Rameshti, F. G. Ghamsari, and R. Asgari, Phys. Rev. B {\bf 95} 045422 (2017).

\bibitem{asgari}
M. Elahi, K. Khaliji, S. M. Tabatabaei, M. Pourfath and R. Asgari, Phys. Rev. B {\bf 91}, 115412 (2015).

\bibitem{landauer}
R. Landauer, IBM J. Res. Dev. {\bf 1}, 223 (1957); {\bf 32}, 306 (1988).

\bibitem{Stiles}
M. D. Stiles, A. Zangwill, Phys. Rev. B {\bf 66}, 014407 (2002).

\bibitem{Born}
M. Born, and E. Wolf, {\it Principles of Optics} (Cambridge University Press, Cambridge, England, 1959).

\bibitem{Majidi11}
L. Majidi, and M. Zareyan, Phys. Rev. B. {\bf 83}, 115422 (2011).

\bibitem{Majidi13}
L. Majidi, and M. Zareyan, J. Comput. Electron. {\bf 12}, 134 (2013).

\bibitem{FS}
J. Linder, A. Brataas, Z. Shomali, and M. Zareyan, Phys. Rev. Lett. {\bf 109}, 237206 (2012).
\end{thebibliography}
\end{document}